# New Insights into Refractive Indices and Birefringence of Undoped and MgO-Doped Lithium Niobate Crystals at High Temperatures

Nina Hong,[1]* Jiarong R. Cui,[2]* Hyun Jung Kim,[3] Ross G. Shaffer,[2] and Nguyen Q. Vinh[2†]

[1] J.A. Woollam Co., Inc. 311 South 7th St. Lincoln, NE 68508 United States

[2] Department of Physics and Center for Soft Matter and Biological Physics, Virginia Tech, Blacksburg, Virginia 24061, United States

[3] NASA Langley Research Center, 1 NASA Driver, Hampton, VA 23681 United States

* Contributed equally to this work

† Corresponding author: email: vinh@vt.edu, phone: 1-540-231-3158

**Abstract**

The lithium niobate single crystal is a well-known optical material that has been employed in a wide range of photonic applications. To realize further applications of the crystal, the birefringence properties need to be determined over a large range of temperatures. We report refractive indices and birefringence properties of undoped and MgO-doped lithium niobate crystals with high accuracy using spectroscopic ellipsometry in the spectral range from 450 to 1700 nm and a temperature range from ambient temperature to 1000 °C. The birefringence results indicate a transition temperature, where the crystal transforms from an anisotropic to isotropic property, and the advance of MgO doping in the crystal, which is related to the optical damage threshold of the materials. In addition, the lattice dynamics of the crystals have been analyzed by revisiting the Raman spectroscopy. The results establish the foundation of optical properties of lithium niobate crystals, providing pathways for their photonic applications.

**Keywords:** Birefringence, Lithium Niobate, MgO-doped Lithium Niobate, Ellipsometry, Raman spectra

## 1. INTRODUCTION

Lithium niobate ($LiNbO_3$) crystals perform multiple functions[1-3] including piezoelectric,[3] ferroelectric,[1, 4] electro-optic,[5] pyroelectric,[6, 7] and nonlinear optical behaviors.[8-11] Specifically, $LiNbO_3$ crystals belong to the *R3c* space group, and below the Curie temperature the crystal shows spontaneous polarization, leading to piezoelectric and ferroelectric properties.[3, 8] Thus, $LiNbO_3$ has received much attention for many technological applications in optical modulators,[5] optical lens,[6] surface acoustic wave devices and sensors,[12] lasers, photodetectors, second-harmonic generators,[13, 14] quasi-phase matching devices,[15] and quantum-related solid-state physics.[1, 2] The crystal belongs to a stable group of optical materials and has a richness of large-magnitude physical effects. Thus,

LiNbO$_3$ has been extensively studied with modifications in growth conditions and investigations on new dopants, since it was grown for the first time.[1, 16]

Smart optical devices using LiNbO$_3$ crystals can be further realized via the precise tunability of refractive properties of the crystals by applying electric fields and/or thermal gradients.[5, 6, 9] These devices include programmable micro-lens based on the micro-zone-plate or photon sieve configuration, which are fabricated on electro-optic LiNbO$_3$ layers to construct tiny spectrometers (i.e., neural proves and on-site soil assay probes, real time analysis of Martian soils or Lunar).[6, 9, 17, 18] To design such devices and predict their functional, thermal and spectral characteristics, both the ordinary, $n_o$, and extraordinary, $n_e$, refractive indices as a function of wavelength and temperature must be accurately determined.[19-22] Moreover, many investigations have suggested that MgO doping at concentrations ~5 mol.% remarkably reduces optical damage in congruent LiNbO$_3$ crystals.[10, 11, 23-26] Therefore, the refractive index as well as birefringence properties of undoped and MgO-doped LiNbO$_3$ crystals need be studied in detail for the optical performance of optical devices.[24, 26] Although several investigations on the refractive index properties of LiNbO$_3$ crystals have been reported, the data obtained so far are not sufficient, given a description of ordinary and extraordinary refractive indices of the crystals in a limited range of temperature and wavelength.[20-22]

In the investigation, we present measurements of the refractive index and birefringence properties of undoped and MgO-doped LiNbO$_3$ single crystals using spectroscopic ellipsometry in a spectral range from 450 to 1700 nm and a temperature range from ambient temperature to 1000 °C. From these samples, a completed refractive index database for both undoped and MgO-doped LiNbO$_3$ crystals is introduced for the first time. The results reveal (*i*) transition temperatures of birefringence, at which the crystals transform from anisotropic to isotropic properties, (*ii*) birefringence trend by doping concentration, which is related to the optical damage threshold, and (*iii*) reversibility of birefringence properties during both heating up and cooling down of the samples. In contrast to previous observations[22] which reported refractive indices ($n_o$ and $n_e$) of the crystals using an interferometer, we determine the values using spectroscopic ellipsometry, a highly accurate and precise measurement technique,[27, 28] and study the lattice dynamics in the crystals with Raman spectroscopy, providing important information of the crystals for photonic applications.

## 2. EXPERIMENTAL METHODS

**2.1. Samples.** During the growing process from melting, LiNbO$_3$ crystals usually render a Li-deficient structure and a lot of defects in the congruent composition, typically, 48.5 mol% Li$_2$O and 51.5 mol% Nb$_2$O$_5$. Thus, a large number of intrinsic defects present in the crystals, produced to maintain the charge neutrality.[29] As a result, the LiNbO$_3$ crystal is a host for incorporation of dopants such as metal ions or rare earth elements with a large concentration. The doping will significantly change the optical properties of LiNbO$_3$ crystals.[1, 30] To explore the optical properties of doped LiNbO$_3$ crystals over a large range of temperatures, we study congruent undoped and MgO-doped LiNbO$_3$ single-crystal wafers. These wafers were purchased from Precision Micro-Optics Inc. MA, USA. The wafers were 3 inches with *z*-cut optical grade. The doped wafers have doping concentrations of 4.7 and 5.5 mol.% of MgO. The doping concentration of Mg in LiNbO$_3$ materials has been determined by several methods during the growth from a melt and examined after the growing process. Two samples with different concentrations have been chosen to explore the photorefractive threshold of the doping level in LiNbO$_3$ crystals. As reported in the literature, the doping level of Mg above the threshold concentration (at ~5 mol.% of MgO in a congruent LiNbO$_3$ crystal) can significantly improve the optical damage resistance of the crystals.[29] Above the threshold concentration, an abrupt change of physical properties of the crystals has also been reported by a number of experiments, including Curie temperature,[31] infrared (IR),[29, 32] ultra-violet,[33] Raman spectra.[34]. To explore the threshold concentration of Mg doping in LiNbO$_3$ crystal on the optical damage resistance, we have performed infrared absorption spectra of Mg-doped LiNbO$_3$ materials. The measurements have been done using an FT-IR spectrometer (Nicolet iS50). The spectrometer has a spectral range from 15 to 6,000 cm$^{-1}$ and the spectral resolution better than 0.09 cm$^{-1}$. Absorption spectra of Mg-doped LiNbO$_3$ materials show two absorption peaks at 3485 and 3535 cm$^{-1}$ (2.87 and 2.83 µm), respectively, (Figure S1, Supplementary Material). The IR absorption spectra show one broad peak for the undoped sample, two absorption lines at 3485 cm$^{-1}$ (2.87 µm) and 3535 cm$^{-1}$ (2.83 µm) for the 4.7 mol.%, and one peak appearing at 3535 cm$^{-1}$ for the 5.5 mol.% of MgO-doped LiNbO$_3$ samples. The absorbance of these bands has been reported in the literature.[29, 35] Specifically, the absorption band at 3485 cm$^{-1}$ (2.87 µm) is related to the OH$^-$ stretching vibration for protons occupying intrinsic $V_{Li}^-$ defects near $Nb_{Li}^{5+}$ (an Li$^+$ ion substituted by an Nb$^{5+}$ ion) in congruent MgO-doped LiNbO$_3$ crystals.[35] The absorption band is absent for Mg doping concentrations higher than 5 mol.%.[29, 32] However, the absorption band at 3535 cm$^{-1}$ (2.83 µm) arises for doping concentration of 4.5 mol.% and above. The absorption spectra indicate a change in the physical properties of the crystals. The doping level in the 5.5 mol.% MgO-doped LiNbO$_3$ sample is above the threshold concentration for the optical damage resistance, while the Mg concentration in the 4.7 mol.% MgO-doped LiNbO$_3$ sample is just at the threshold.[29, 32] To understand the lattice structure, impurities and defects of these LiNbO$_3$ crystals, we have employed Raman spectroscopy and spectroscopic ellipsometry to study the optical properties of the crystals. Raman spectra provide information on the Li-deficient structure and defects in undoped as well as MgO-doped crystals. The spectroscopic ellipsometry has been employed to evaluate birefringence properties of these crystals over a large range of temperatures with different doping levels.

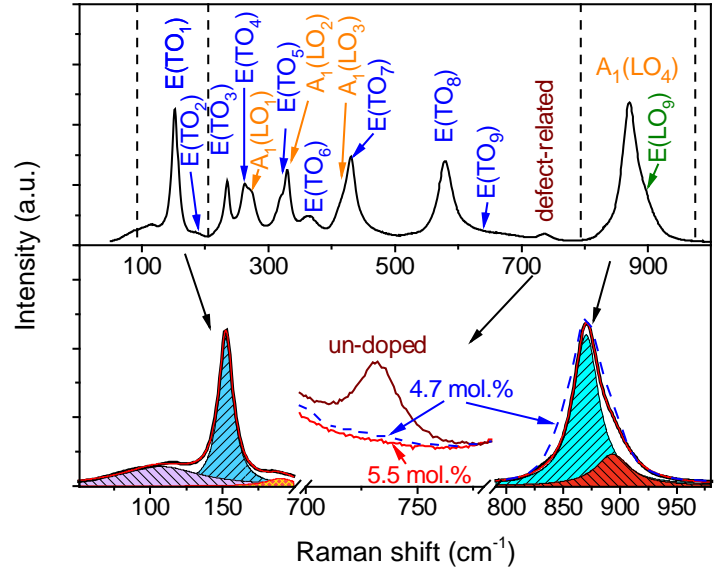

**Figure 1:** Raman spectra recorded in the z-direction for undoped and Mg-doped LiNbO$_3$ crystals. (top) Raman modes of E(TO) and A$_1$(LO) lines have been observed, including 9 E(TO), 4 A$_1$(LO) and defect-related lines. (bottom) The Raman E(TO$_1$) mode is decomposed into three components and the A$_1$(LO$_4$) mode is analyzed into two components. A defect-related mode has been observed at 737 cm$^{-1}$ only for undoped crystals. The Raman A$_1$(LO$_4$) phonon mode is broader at doped crystals.

**2.2. Raman spectroscopy.** Raman spectroscopy has been widely used for studying the lattice and chemical structure, impurities and defects, phase transitions and phonon dynamics of crystals based on the interaction of light with chemical bonds within the crystals. At room temperature, LiNbO$_3$ crystals belong to the *R3c* space group, and thus, 18 vibrational modes at zero phonon are decomposed into 4A$_1$ + 9E + 5A$_2$. The A$_1$ and E phonons are infrared (IR) and Raman active, but A$_2$ modes are Raman inactive. A large number of experimental and theoretical investigations have been reported in the literature; however, the assignment of E phonon modes is not yet unambiguously established and some controversies remain between authors due to the presence of both intrinsic and extrinsic defects in LiNbO$_3$ crystals.[36-38] To revisit the Raman spectroscopy, spectra have been recorded with a Raman microscope, model Horiba XploRA PLUS operated with a movable and high groove density grating (2400 grooves/mm). The samples are illuminated by a green laser at 532 nm. An Olympus microscope (BX41) with 100× objective lens is used for focusing the laser on the sample and collecting scattered light in a 180º backscattering configuration. The Raman spectra were averaged from 3 scans in which each scan is accumulated over 5 seconds. The Raman spectra of LiNbO$_3$ crystals are plotted in Figure 1.

**2.3. Spectroscopic ellipsometry.** Spectroscopic ellipsometry is a powerful technique employed to characterize light reflection (or transmission) from samples, allowing the evaluation of optical constants, including extinction coefficient, $k$, and refractive index, $n$, and thickness of thin films. The technique measures directly the change in the polarization state of a polarized light after it reflects from the film surface, including the amplitude ratio and phase difference. The ratio, $\rho$, of the amplitude reflection coefficient for $p$- and $s$-polarizations is given by:

$$\rho = \frac{r_p}{r_s} = \tan\Psi \, e^{i\Delta} \qquad (1)$$

where $\Psi$ represents the angle determined from the amplitude ratio, $\Delta$ expresses the phase difference between the $s$- and $p$- polarized light, $r_s$ and $r_p$ are the perpendicular and parallel components, respectively, of the reflected light with respect to the plane of incidence.[27, 28] Spectroscopic ellipsometry offers high-precision measurements for ($\Psi$, $\Delta$), optical constants, and film thickness of materials. The technique determines relative light intensities modulated by optical elements, instead of the absolute light intensities of reflected $s$- and $p$-polarizations. If we compare the values with absolute reflectance measurements, measurement errors induced by various imperfections in optical instruments become very small in ellipsometry measurements. Consequently, using the ellipsometry technique, optical constants and film thicknesses can be determined with high precision.

The ellipsometry data of optical grade $z$-cut LiNbO$_3$ single crystals as a function of temperature were measured in a wide spectral range from 190 nm to 1700 nm using the J.A. Woollam RC2 spectroscopic ellipsometer. The RC2 spectroscopic ellipsometer features innovative new technology, including a high spectral resolution of 1.0 nm at $\lambda$ < 1000 nm and 2.5 nm at $\lambda$ > 1000 nm, achromatic compensator design, and dual rotating compensators. The optical design allows exceptional data accuracy for spectroscopic ellipsometry measurements The sample temperatures were controlled from ambient to 1000 °C with a stability of < 1 ºC using a Linkam TSEL 1000 heat stage. While the maximum ramp rate of the heat stage is 200 ºC/minute, we increased (or decreased) the temperature in steps of 25 ºC (or 50 ºC) and waited for a few minutes before each ellipsometry measurement to obtain temperature stability.

The $z$-cut LiNbO$_3$ crystals are uniaxially anisotropic with the optic axis parallel to the sample normal direction. We measured both double-side polished and single-side polished samples at three different angles (50, 55, and 60 degrees) to build a robust uniaxial model. Measurement results of double-side polished samples have provided excellent sensitivity to the birefringence, $\Delta n = n_e - n_o$, due to their long optical path length, leading to high-frequency interference oscillations caused by the difference between refractive indices of the in-plane (ordinary) and out-of-plane (extraordinary) directions. Measurements from single-side polished samples have enhanced the sensitivity of the absolute refractive index of the LiNbO$_3$ samples.

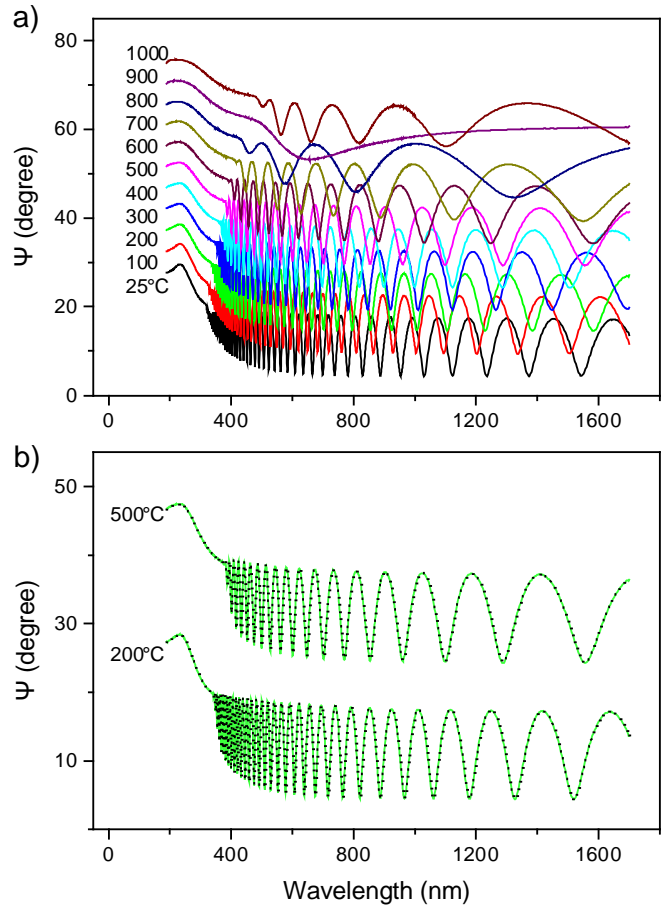

**Figure 2**: Spectroscopic ellipsometry data for the amplitude ratio in degree, Ψ, of the *z*-cut undoped LiNbO$_3$ crystal at elevated temperatures. (top) Spectra have been collected from 193 nm to 1700 nm with temperature varied from 25 ºC to 1000 ºC. An offset of 5 degrees is applied to separate neighboring spectra. (bottom) The spectra show a reversibility during the heating up (solid line) and cooling down (dotted line) of the temperature.at 250 ºC and 500 ºC.

## RESULTS AND DISCUSSION

Raman spectra of our samples have been obtained in the parallel $z(xx)\bar{z}$ polarization geometry for which E(TO) and A$_1$(LO) phonon modes are expected.[39] Carefully analyzing Raman spectra of our samples, we have observed eleven optical phonons that are assigned as 9 E(TO) lines and 4 A$_1$(LO) modes (Figure 1). Four A$_1$(LO) optical phonons have been identified in the Raman spectrum at 272.6, 330.0, 418.0, and 871.1 cm$^{-1}$. These values are widely accepted by most authors.[36-38, 40] For the undoped LiNbO$_3$ crystal, nine TO optical phonons of the E-symmetry have been observed at 152.3, 183.8, 235.8, 262.5, 321.0, 364.0, 430.3, 579.2, 670 cm$^{-1}$. A phonon line at ~737 cm$^{-1}$ has been under discussion.

Some authors recently assigned this Raman mode to an optical phonon mode.[36, 39] However, this mode disappears and reappears in doped LiNbO$_3$ crystals depending on the doping concentration and composition of the materials. Thus, this phonon mode could be related to a defect vibration mode and affected to optical properties of the crystals.

The Raman measurements are extremely sensitive to small modifications of crystals such as growth conditions including congruent and stoichiometry compositions (Li/Nb ratio),[37] dopants (e.g. rare-earth elements, transition metals), doping concentration and defects[41, 42] as well as optical configurations for the measurements.[36, 39, 43] Particularly, congruent LiNbO$_3$ crystals usually present Li deficiency, and thus, they contain a considerably large amount of intrinsic defects.[22] The deficit of Li ions in non-stoichiometric LiNbO$_3$ crystals is expected to be compensated by Nb ions on Li lattice sites, forming Nb anti-site defects.[44] An addition of Li into LiNbO$_3$ crystals modifies the crystal structure, and thus, varies the macroscopic dielectric property of the crystals. The Li deficit shows up as a linear behavior in the variation of the dielectric tensor.[38] Such linear behavior indeed has been found in the linear narrowing of the Raman linewidth when the composition of LiNbO$_3$ changes from congruent to stoichiometric crystals.[37, 45] Specifically, the linewidths of strong intensity bands including E(TO$_1$), A$_1$(TO$_1$), A$_1$(LO$_4$), usually vary with the Li/Nb ratio in undoped crystals, thus, the information of some Raman modes can be employed to determine the composition of LiNbO$_3$ crystals.[37, 39, 46] The Raman linewidth for the lowest-frequency E(TO$_1$) reduces significantly from ~10.4 cm$^{-1}$ to 7 cm$^{-1}$, for a congruent sample and nearly stoichiometric material, respectively,[37] corresponding to the composition, $x_c = [Li]/([Li] + [Nb])$, of 48.5% to 49.74%, respectively. The composition, $x_c$, of LiNbO$_3$ crystals can be determined from the linewidths of some Raman modes, as given by Schlarb *et al*.[39, 46] Linear relations were established to estimate the Li concentration ($C_{Li}$, mol.%) as a function of Raman linewidths ($\Gamma$, cm$^{-1}$) in undoped LiNbO$_3$ crystals,[46] $C_{Li} = 53.03 - 0.4739\,\Gamma_{\text{undoped}}$ for the E(TO$_1$) phonon at 152.3 cm$^{-1}$ and $C_{Li} = 53.29 - 0.1837\,\Gamma_{\text{undoped}}$ for the A$_1$(LO$_4$) phonon at 871.1 cm$^{-1}$. Note that some Raman modes contain weak contributions from a mixing of the mode with other Raman modes, a slight mismatch of the crystal orientation or a macroscopic damage of the wafer. To determine linewidths of these modes, we make a deconvolution of the Raman modes into several components with Lorentzian functions (Figure 1, bottom).[47] We have employed the method and estimated the composition, $x_c$, of our undoped samples based on the full width at half maximum of the two Raman modes. The composition, $x_c$, of 48.2 ± 0.4% is an average of estimations of 47.8% and 48.5% from E(TO$_1$) and A$_1$(LO$_4$) Raman modes, respectively, in our Raman spectrum.

By incorporating dopants such as Mg, Zn, In or Sc in the congruent LiNbO$_3$ lattice, optical properties including birefringence and optical damage threshold have been improved.[48] Among them, Mg is the most frequently used, and accordingly, the optical damage is strongly reduced as well as the birefringence is enhanced by the introduction of 4.6 mol.% Mg in a congruent melt.[48, 49] The dopants significantly affect phonon modes recorded in Raman spectra. The linewidth of Raman modes of MgO-doped LiNbO$_3$ crystals is broader when the Mg doping concentration increases.[50, 51] The positions of these Raman modes also change.[51] Nataf *et al*.,[50] estimated the linewidth as a function of Mg concentration in LiNbO$_3$ crystals for E(TO$_1$), E(TO$_8$) and A$_1$(LO$_4$) Raman modes with the slopes of 0.24, 0.64, and 1.31, respectively. We have employed the parameters into our MgO-doped LiNbO$_3$ samples. Indeed, we have obtained similar values of Mg concentrations as indicated in our wafers. Concurrently, the increase of linewidth of Raman modes at E(TO$_1$) and A$_1$(LO$_4$) from undoped to 4.7 and 5.5 mol.% samples provides information on the Li content in the crystals. By employing the linewidth estimation method, we have determined the Li concentration, $x_c$, of ~ 46.8% for two doped samples.

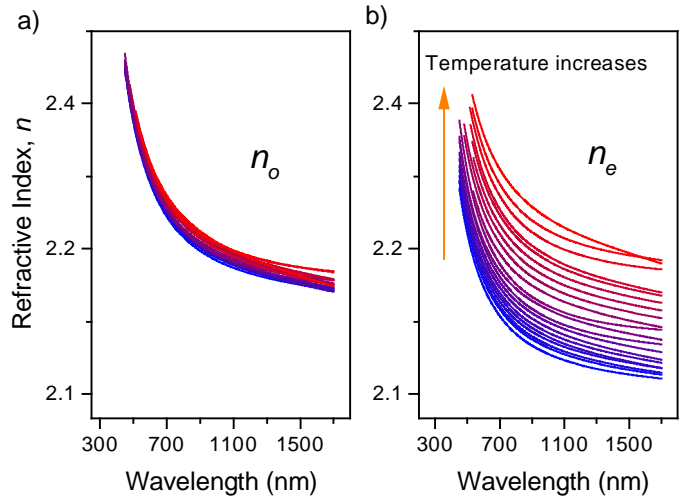

**Figure 3**: Spectroscopic refractive indices of the *z*-cut undoped LiNbO$_3$ crystal in the sample plane (a) and normal (b) directions. The measurement temperature varied from 25 ºC to 1000 ºC with a step of 25 ºC is scaled by the color gradient from blue to red.

The optical properties of doped crystals are linked to the incorporation site of Mg ions in the LiNbO$_3$ lattice. Several calculations and experiment results showed that, at first, Mg ions replace the Nb anti-sites in congruent crystals, and further, Mg ions are incorporated into Li sites at a larger concentration.[11, 42] A small Raman peak at ~737 cm$^{-1}$ observed in the $z(xx)\bar{z}$ backscattering geometry has been assigned earlier to ilmenite-like stacking defect vibration[52] and related to the incorporation of Mg ions into LiNbO$_3$ crystals. The Raman intensity of the defect band at ~737 cm$^{-1}$ obviously can be seen in our undoped crystal (Figure 1, bottom), while this peak in congruent MgO-doped samples (~5 mol.%) disappears. However, the peak has been reported in congruent samples with a weak intensity at higher MgO-doped concentration of 7.8 mol.%, and more intensely in stoichiometric LiNbO$_3$:Mg samples.[43, 52] The dependence of this peak on Li content and MgO doping concentration in the crystals suggested a defect structure which has ilmenite-like stacking and a direct relationship with anti-sites of $Nb_{Li}$ ions. When Mg dopant is introduced into the $LiNbO_3$ lattice, the stress forming from the ilmenite-like stacking reduces, thus, the intensity of the Raman peak becomes weaker. However, when the Mg concentration in the crystal exceeds the threshold of the stress produced from impurity defects, the Raman mode reappears. This defect is in good agreement with the Li vacancy model.[53] The results of impurity and defect concentration in the crystals have provided useful information to understand birefringence properties of the crystals using spectroscopic ellipsometry.

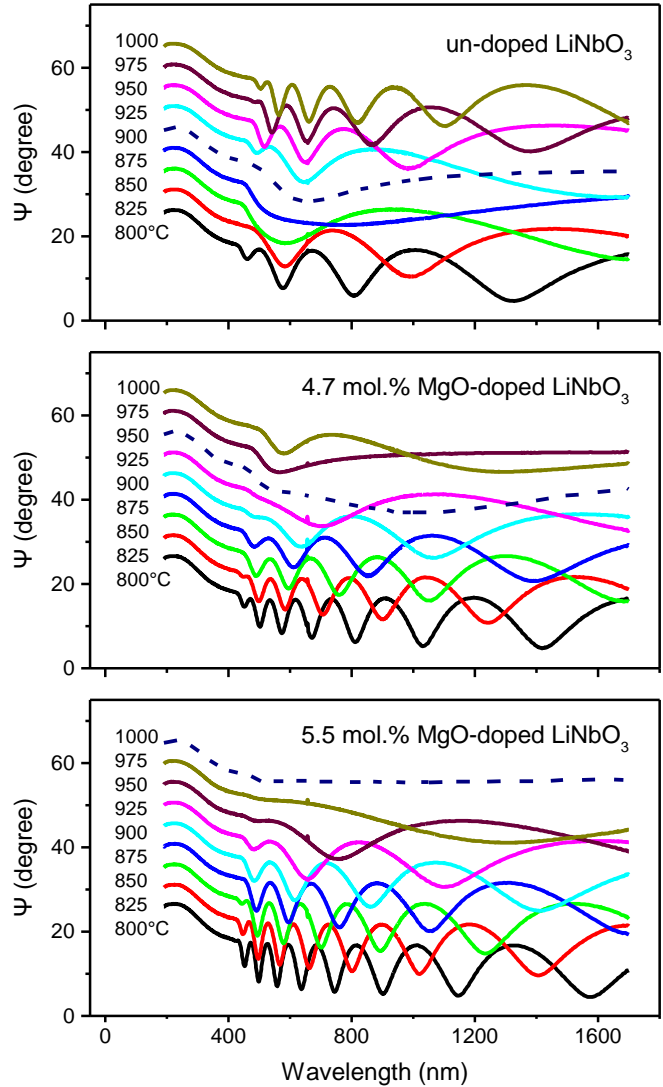

**Figure 4**: Spectroscopic ellipsometry data, $\Psi$, of $z$-cut undoped, 4.7 and 5.5 mol.% MgO-doped $LiNbO_3$ crystals at temperatures from 800 ºC to 1000 ºC. The dotted lines indicate the transformation of the crystals from anisotropic to isotropic properties, where the inversion of birefringence sign occurs.

Spectroscopic ellipsometry has provided excellent precision measurements of refractive index and birefringence properties, $\Delta n$, of materials.[28] The amplitude ratio in degrees, $\Psi$, of ellipsometry data of the undoped double-side polished LiNbO$_3$ crystal has been collected from 190 to 1700 nm at elevated temperatures from 25 to 1000 ºC (Figure 2). To separate the $\Psi$ curves, an offset of 5 degrees is applied between neighboring spectra. The spectroscopic ellipsometry data consist of two regions separated by the bandgap energy of the LiNbO$_3$ crystal. Below the bandgap energy, the LiNbO$_3$ crystal is optically transparent and creates spectral oscillations. The oscillations are caused by the birefringence splitting of light that travels through the substrate and returns to the surface, which makes a long optical path length. Assuming the substrate thickness does not change, the number of oscillations is proportional to the difference of the refractive indices between the ordinary and extraordinary directions. Given the number of oscillations, the data reveal that the refractive index difference reduces as the temperature increases. When the temperature reaches ~900 ºC, no oscillation is found, indicating that the sample becomes optically isotropic. Then, the birefringence effect reappears at higher temperature spectra.

We collected ellipsometry spectra from the three samples during heating up as well as cooling down between ambient temperature and 1000 ºC to test the reversibility of their optical properties. Figure 2b shows a representative data set at 250 and 500 ºC (shifted by 20 degrees) of undoped LiNbO$_3$ crystal. Solid green lines are measured during the heating up and dotted lines are measured during the cooling down sweeps. While not all data are shown here, all samples reveal excellent reversibility of the ellipsometry data.

Above the bandgap energy, the LiNbO$_3$ crystal absorbs light, and the signal comes from the surface only. Therefore, the difference of refractive indices does not create spectral oscillations. We can determine approximate bandgap energy from the spectroscopic ellipsometry data by finding the transition point where the spectral oscillations diminish. Thus, we can conclude that the bandgap shows a red shift as the temperature increases. However, an accurate determination of the bandgap energy associated with extinction coefficients, $k_o$ and $k_e$, of uniaxially anisotropic LiNbO$_3$ crystals requires additional samples where the extraordinary axis is in the sample plane.[27, 28] Also, a number of optical transitions corresponding to various inter-band transitions occur in the crystal, and thus, the modeling of dielectric functions generally becomes complicated, and we do not focus on this discussion in this work.

The data analysis of spectroscopic ellipsometry for transparent semiconductors in the visible region, where the absorption is very low, can be performed rather easily, because dielectric functions commonly exhibit simple structures that can be represented by the Cauchy or Sellmeier model. In this case, the dielectric function of a semiconductor changes smoothly, showing a normal dispersion. A simple optical model including two Lorentzian components corresponding to two optical transitions outside the measured range, one in the ultraviolet (UV) region (an upper energy limit) and another in the IR range (a lower energy limit), has been employed to determine the Kramers-Kronig refractive index as a function of wavelength in the transparent region. Each component has two independent parameters including the energy of the transition and amplitude of the component. The refractive indices in the ordinary and extraordinary directions of the spectroscopic ellipsometry data have been determined using the data analysis software, CompleteEASE, which employs the Levenberg-Marquardt algorithm to obtain the best-fit parameters and minimizes the mean-squared error between the measured and model-generated data. The random noise was estimated 30 times using the Bootstrap Monte-Carlo simulations.[54] The dual-rotating configuration of the ellipsometer minimizes systematic errors as the two compensators are continuously rotating. Thus, the systematic errors were computed assuming a conservative angle offset of 0.02º as the primary source. The random noise of the refractive indices in the ordinary and extraordinary directions ($n_o$ and $n_e$) and the birefringence ($\Delta n = n_e - n_o$) were estimated in the low fourth or high fifth decimal point, leading to very high precision. The systematic errors of the absolute refractive index and birefringence have been determined of ± 0.002 and ± 0.0001, respectively. The birefringence is more accurately determined as it is associated with the number of oscillations in the spectroscopic ellipsometry data rather than their positions.

The refractive indices of the $z$-cut undoped LiNbO$_3$ crystal have been extracted at elevated temperatures from 25 to 1000 ºC with a step of 25 ºC for the ordinary (Figure 3a), and extraordinary (Figure 3b) directions. While the refractive index of the $z$-direction (extraordinary, $n_e$) increases, the refractive index in the sample plane (ordinary, $n_o$) is relatively stable when the temperature increases. This behavior shows good agreement with X-ray diffraction results studied by Abrahams *et al*.[4] This indicates that the birefringence, $\Delta n = n_e - n_o$, of the crystal reduces at high temperatures. The birefringence sign inverts from negative ($n_o > n_e$) to positive values ($n_e > n_o$) when the temperature of the sample is higher than 900 ºC for the undoped LiNbO$_3$ crystal.

To explore the inversion of the sign of birefringence properties, we have performed temperature-dependent ellipsometry experiments with different doping concentrations of MgO in LiNbO$_3$ crystals. The amplitude ratio spectra, Ψ, of ellipsometry measurements of all samples above 800 ºC have been plotted in Figure 4. Again, to separate the Ψ curves, an offset of 5 degrees is applied between neighboring spectra. The dotted lines represent where the minimum of the birefringence occurs, indicating the near isotropic structure of the crystals. The inversion of the birefringence sign occurs near 900 ºC, 950 ºC, and 1000 ºC for undoped, 4.7 and 5.5 mol.% of MgO-doped LiNbO$_3$ crystals, respectively. The increase of the inversion temperature with MgO doping concentration indicates that the dopant level is related to the variation of lattice constants of LiNbO$_3$ with temperature, and the 5.5 mol.% MgO-doped LiNbO$_3$ crystal is the most stable among the three crystals.

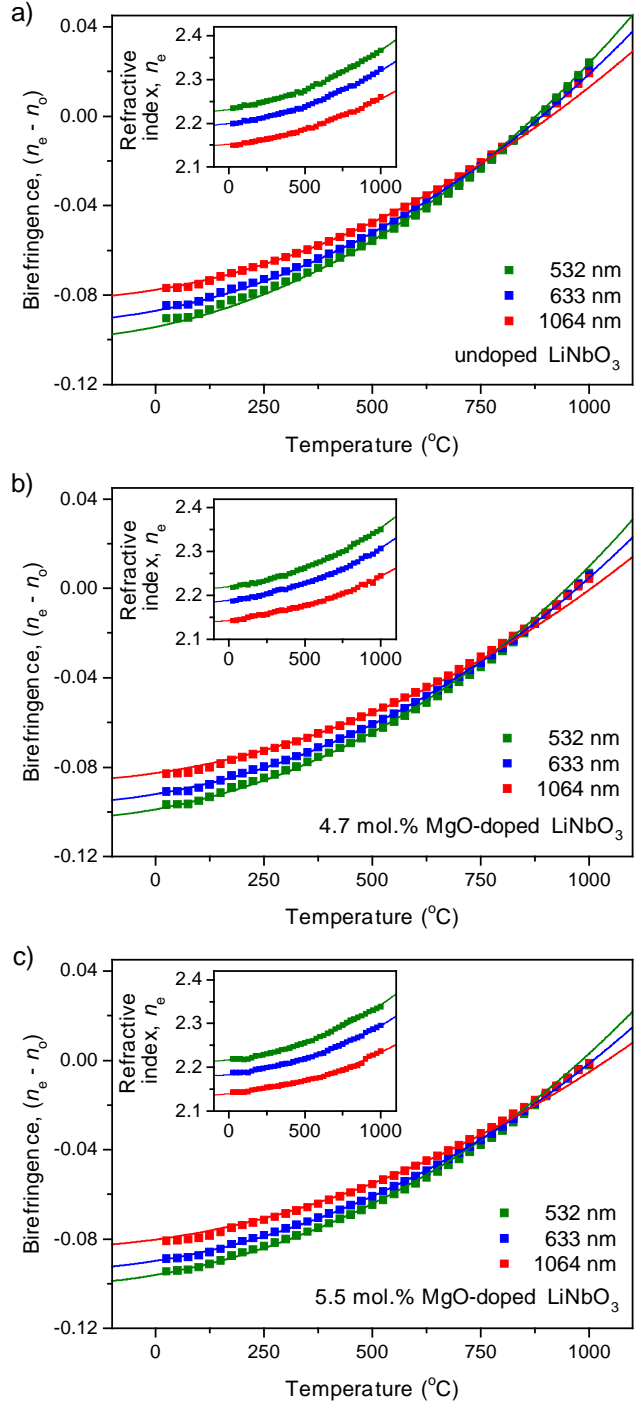

**Figure 5**: Birefringence properties of LiNbO$_3$ crystals as a function of temperature for λ = 532, 633, 1064 nm. The difference of $n_e$ and $n_o$ for all wavelengths studied here reduces when temperature increases for (a) undoped, (b) 4.7 and (c) 5.5 mol% MgO-doped LiNbO$_3$. Solid curves are the best fits of the experimental data to the five-oscillator Sellmeier equation (Eq. 3). Insets present extraordinary refractive indices ($n_e$) of the LiNbO$_3$ materials for the three wavelengths at different temperatures. Error bars are smaller than the symbol size.

To determine the interaction of electromagnetic waves and LiNbO$_3$ crystals as a function of temperature, multi-oscillators for the refractive indices of the crystals have been reported in the literature.[22, 49, 55-57] A common approach to describe the dispersion of the refractive index is a sum of multiple Sellmeier components,[46]

$$n^2 = 1 + \sum_j \frac{A_j}{\lambda_j^2 - \lambda^2} \quad (2)$$

where $\lambda_j$ is the resonant wavelength of the $j^{th}$ oscillator, and $A_j$ is proportional to the transition probability for the optical excitation and number of oscillators per volume. A number of models have been provided to explain the refractive index of LiNbO$_3$. Uchida[56] proposed a Sellmeier model, which contains two UV oscillators of oxygen 2p orbitals forming a filled valence band and the lowest lying conduction band located at 5 eV (248 nm) and 9 eV (138 nm). Schlarb *et al.*,[55, 56] suggested four oscillators including two adjacent vibrational modes located at around 5 eV (248 nm), a plasmon at far UV of ~7 eV (177 nm) and one for the mid-infrared (mid-IR) term with the long wavelength approximation. Zelmon *et al.*,[58] employed a three-oscillator Sellmeier equation with two UV terms at ~5 eV (248 nm) and 9 eV (138 nm) as predicted by Uchida and Mamedov,[55, 56] and one IR oscillator. By employing a wide temperature range, these models are not sufficient to reproduce our results. We have employed a five-oscillator Sellmeier equation with one plasmon at the far UV with energy > 6.2 eV (200 nm), two UV and two IR oscillators.

$$n^2(T) = 1 + A_{\text{plasmon}} + \frac{50 + c_{\text{Li}}}{100} \frac{A_{\text{UV1}}}{\lambda_1^{-2} - \lambda^{-2}} + \frac{50 - c_{\text{Li}}}{100} \frac{A_{\text{UV2}}}{\lambda_2^{-2} - \lambda^{-2}} + \frac{A_{\text{IR}}}{\lambda_3^{-2} - \lambda^{-2}} - A_{\text{mid-IR}} \lambda^2 \quad (3)$$

where $c_{\text{Li}}$ is the Li concentration determined from the linewidth of two phonon modes, E(TO$_1$) and A$_1$(LO$_4$), in the Raman spectrum, $A_{\text{plasmon}}, A_{\text{UV1}}, A_{\text{UV2}}, A_{\text{IR}}, A_{\text{mid-IR}}$ are contributions to the refractive index from plasmons, two oscillators in UV region, one oscillator in IR range and mid-IR term. These parameters are assumed to be concentration dependent of dopants as well as the sample temperature. The first term represents a contribution to the refractive index from the far UV region, in which the resonant wavelength is much shorter that the excitation wavelength ($\lambda_{\text{plasmon}} \ll \lambda$), thus, the Sellmeier approximation for plasmons ($A_{\text{plasmon}} \approx 1 + A\lambda_{\text{plasmon}}^2$) can be estimated as a constant.[22] Three Sellmeier oscillators stand for contributions of oscillators which happened in the UV and IR regions. The last term describes contributions from mid-IR oscillators. Due to long wavelengths from these oscillators ($\lambda_{\text{oscillators}} \gg \lambda$), the Sellmeier terms can be approximated as $-A_{\text{mid-IR}} \lambda^2$.

The wavelengths of resonant oscillations are sensitive to temperature. Effectively, the oscillation wavelengths account for thermal effects as $\lambda_i(T) = \lambda_{\text{osc\_i}} + B_i f$, where $\lambda_{\text{osc\_i}}$ is the vacuum wavelength of the oscillator, $B_i$ is a constant, and *f* is a temperature dependence parameter given by:[20]

$$f = (T - T_0)(T + T_0 + 2 \times 273.16) \quad (4)$$

where $T_0$ is a constant (24.5 °C) and $T$ is the temperature of the sample in degree centigrade. When $T = T_0$, Eq. 4 reduces to the temperature-independent Sellmeier equation.

**Table 1.** Parameters have been obtained from fitting experimental results to Eq. 3, containing multiple Sellmeier components.

| | un-doped | | MgO-doped with 4.7-mol. % | | MgO-doped with 5.5-mol. % | |
|---|---|---|---|---|---|---|
| | $n_o$ | $n_e$ | $n_o$ | $n_e$ | $n_o$ | $n_e$ |
| $A_{plasmons}$ | 2.42 | 2.42 | 2.59 | 2.59 | 2.46 | 2.46 |
| $A_{UV1}$ | $4.91 \times 10^{-5}$ | $4.59 \times 10^{-5}$ | $4.63 \times 10^{-5}$ | $3.97 \times 10^{-5}$ | $4.79 \times 10^{-5}$ | $4.11 \times 10^{-5}$ |
| $\lambda_1$ (nm) | $225 \pm 5$ | $218 \pm 5$ | $225 \pm 5$ | $218 \pm 5$ | $225 \pm 5$ | $218 \pm 5$ |
| $B_1$ | $1.97 \times 10^{-6}$ | $1.35 \times 10^{-5}$ | $3.23 \times 10^{-6}$ | $1.67 \times 10^{-5}$ | $3.13 \times 10^{-6}$ | $1.56 \times 10^{-5}$ |
| $A_{UV2}$ | $1.59 \times 10^{-5}$ | $3.53 \times 10^{-5}$ | $1.18 \times 10^{-5}$ | $6.11 \times 10^{-5}$ | $1.08 \times 10^{-5}$ | $7.19 \times 10^{-5}$ |
| $\lambda_2$ (nm) | 260 | 250 | 260 | 250 | 260 | 250 |
| $B_2$ | $-8.24 \times 10^{-5}$ | $-2.35 \times 10^{-5}$ | $-1.33 \times 10^{-4}$ | $-4.21 \times 10^{-5}$ | $-1.82 \times 10^{-4}$ | $-4.43 \times 10^{-5}$ |
| $A_{IR}$ | $1.32 \times 10^{-8}$ | $9.25 \times 10^{-9}$ | $7.98 \times 10^{-9}$ | $8.06 \times 10^{-9}$ | $1.54 \times 10^{-8}$ | $1.79 \times 10^{-9}$ |
| $\lambda_3$ (nm) | 2900 | 2900 | 2900 | 2900 | 2900 | 2900 |
| $B_3$ | $9.83 \times 10^{-5}$ | $9.83 \times 10^{-5}$ | $9.83 \times 10^{-5}$ | $9.83 \times 10^{-5}$ | $9.83 \times 10^{-5}$ | $9.83 \times 10^{-5}$ |
| $A_{mid-IR}$ | $2.02 \times 10^{-9}$ | $6.52 \times 10^{-9}$ | $2.39 \times 10^{-8}$ | $1.61 \times 10^{-8}$ | $1.96 \times 10^{-8}$ | $6.49 \times 10^{-9}$ |

The high precision data we present here over a large range of temperatures provide new insights into the birefringence properties of the LiNbO$_3$ materials. For example, previous studies of the material to explore the optical properties up to 360 °C focused on oscillators at the UV region and long wavelength in the IR region with an approximation term of $-A\lambda^2$.[20, 22, 49, 56] Zelmon *et al*.,[58] employed three oscillators, including two UV and one IR term, without an oscillation at the long wavelength. In our fits, the eleven parameters in the five-Sellmeier model are varied simultaneously. The fit parameters are provided in Table 1 for undoped congruent, 4.7 and 5.5 mol.% MgO-doped materials. The fit that we have obtained predicts the data within experimental error throughout the temperature range. The wavelengths of the UV oscillators obtained by the fitting routine are consistent with transitions observed in previous reports.[22, 49, 56] In the IR region, the oscillator wavelengths obtained from fitting processes are located at the IR transmission edge of the material.[35, 59] We have obtained one unique set of parameters for all samples using in this research. The best-fit values for the refractive indices, $n_e$ and $n_o$, are included in Table 1 and plotted in Figure 5 (solid curves).

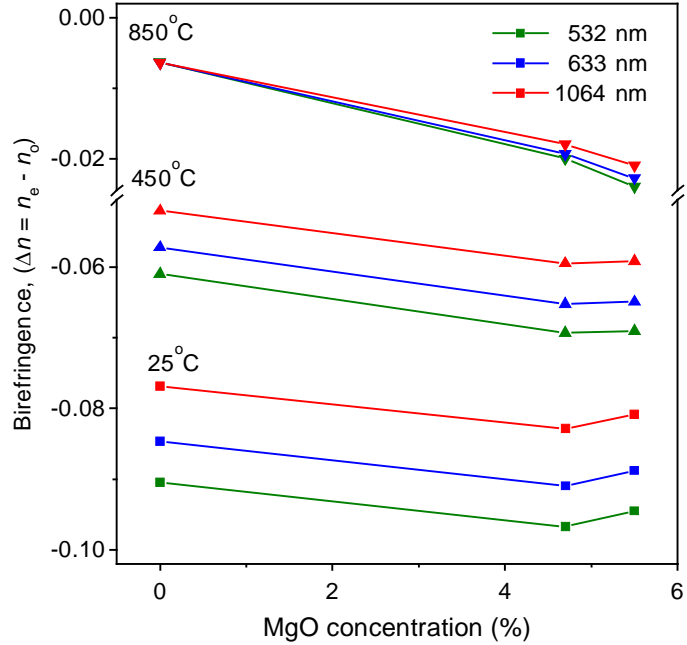

**Figure 6**: Birefringence properties of LiNbO$_3$ crystals as a function of doping concentration for three wavelengths of 532 (green), 633 (blue), 1064 nm (red), at different temperatures. When the doping concentration increases, the difference between $n_e$ and $n_o$ increases and saturates at high doping concentration of ~ 5 mol.%. Error bars are smaller than the symbol size.

To explore the birefringence properties of LiNbO$_3$ materials, we have extracted the birefringence, $\Delta n = n_e - n_o$, as a function of doping concentration from our results. Birefringence properties of LiNbO$_3$ crystals as a function of MgO doping concentration are plotted for three wavelengths of 532 (green), 633 (blue), 1064 nm (red), at different temperatures (Figure 6). When the doping concentration increases, the difference between $n_e$ and $n_o$ increases and saturates at a high doping concentration of ~5 mol.%. Comparing the birefringence results with the defect oscillation mode at ~737 cm$^{-1}$ in the Raman spectra indicates a similar behavior. When the MgO concentration in the material increases, the intensity of the defect Raman mode at ~737 cm$^{-1}$ reduces and disappears, as reported also in the literature.[43, 52]

**CONCLUSIONS**

We have performed Raman and spectroscopic ellipsometry experiments for undoped and MgO-doped LiNbO$_3$ crystals. By employing Raman spectroscopy, we reevaluate and compare the lattice dynamics of these crystals. The results allow us to estimate the deficiency and imperfection of congruent LiNbO$_3$ crystals with and without MgO doping. Using the spectroscopic ellipsometry, we report refractive indices and birefringence properties of undoped and MgO-doped LiNbO$_3$ crystals with high accuracy in the spectral range from 450 to 1700 nm and a temperature range from ambient temperature to 1000 °C. The birefringence results indicate a transition temperature, where the crystal transforms from an anisotropic to isotropic property, and the advance of MgO doping in the crystal, which is related to the optical damage threshold of the materials. The results provide information for characterization and development of optical devices based on lithium niobate single crystals for photonic applications.

**ACKNOWLEDGMENTS**

This work was made possible by the use of Virginia Tech's Materials Characterization Facility, which is supported by the Institute for Critical Technology and Applied Science, the Macromolecules Innovation Institute, and the Office of the Vice President for Research and Innovation. We acknowledge Advanced Research Computing at Virginia Tech for providing computational resources and technical support that have contributed to the results reported within this paper.

# Supplementary Material

# New Insights into Refractive Indices and Birefringence of Undoped and MgO-Doped Lithium Niobate Crystals at High Temperatures

Nina Hong,[1]* Jiarong R. Cui,[2]* Hyun Jung Kim,[3] Ross G. Shaffer,[2] and Nguyen Q. Vinh[2]†

[1] J.A. Woollam Co., Inc. 311 South 7th St. Lincoln, NE 68508 United States
[2] Department of Physics and Center for Soft Matter and Biological Physics, Virginia Tech, Blacksburg, Virginia 24061, United States
[3] NASA Langley Research Center, 1 NASA Driver, Hampton, VA 23681 United States
* Contributed equally to this work
† Corresponding author: email: vinh@vt.edu, phone: 1-540-231-3158

The doping concentration of Mg in lithium niobate ($LiNbO_3$) materials has been determined by several methods during the growth from a melt and examined after the growing process. To explore the effect of doping concentration on the optical damage resistance, we have performed absorption spectra of Mg-doped $LiNbO_3$ crystals. The measurements have been done using an FT-IR spectrometer (Nicolet iS50). The spectrometer has a spectral range from 15 to 6,000 $cm^{-1}$ and the spectral resolution better than 0.09 $cm^{-1}$. Figure S1 shows absorption spectra of undoped, 4.7 and 5.5 mol.% of MgO-doped $LiNbO_3$ samples. The IR absorption spectra show one broad peak for the undoped sample, two absorption lines at 3485 $cm^{-1}$ (2.87 µm) and 3535 $cm^{-1}$ (2.83 µm) for the 4.7 mol.%, and one peak appearing at 3535 $cm^{-1}$ for the 5.5 mol.% of MgO-doped $LiNbO_3$ samples. The absorbance of these bands has been reported in the literature. Specifically, the absorption band at 3485 $cm^{-1}$ (2.87 µm) is absent for Mg doping concentrations higher than 5 mol.%. The absorption band at 3535 $cm^{-1}$ (2.83 µm) arises for doping concentration of 4.5 mol.% and above.[1, 2] The IR absorption spectra indicate a change of physical properties in the crystals. The doping level in the 5.5 mol.% MgO-doped $LiNbO_3$ sample is above the threshold of the optical damage resistance, while the Mg concentration in the 4.7 mol.% MgO-doped $LiNbO_3$ sample is just at the threshold. The observation of the threshold of the optical damage resistance was reported by a number of experiments, including Curie temperature,[3] infrared (IR),[4, 5] ultra-violet,[6] Raman spectra.[7]

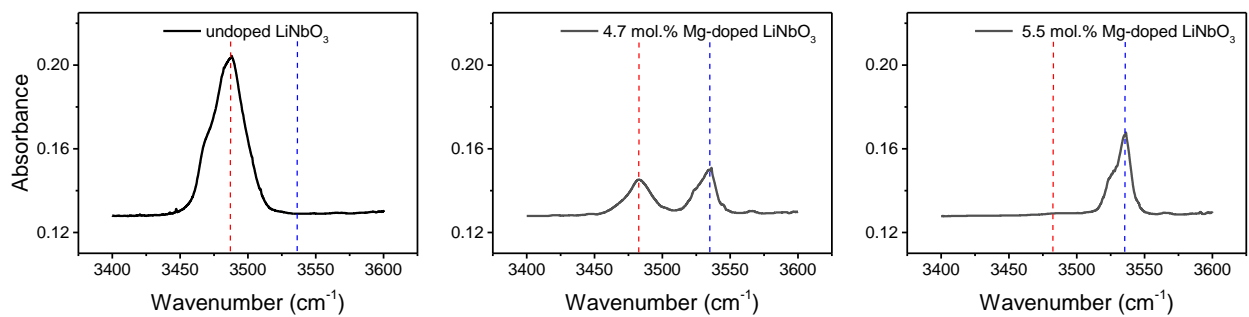

**Figure S1:** Absorption spectra of undoped, 4.7 and 5.5 mol.% of Mg-doped $LiNbO_3$ samples.